\documentstyle[aps,prl,twocolumn,psfig,epsf]{revtex}
\begin{document}
\draft
\twocolumn[
\author{C. Cabrillo, J. L. Rold{\'a}n, P. Garc{\'\i}a-Fernandez}
\address{Instituto de Estructura de la Materia, CSIC,
Serrano 123, 28006 Madrid, Spain.}

\date{\today}
         
\title{Almost perfect squeezing at high intensities by stabilization
from competing non-linearities}

\maketitle
\mediumtext

\begin{abstract}
\leftskip 2.0truecm      % for centering
\rightskip -2.0truecm    % the abstract
\indent
The squeezing properties of a cavity Second Harmonic Generation (SHG) system 
with an added Kerr effect-like nonlinearity are studied as a
function of the intra-cavity photon number.  The competition between the 
$\chi^{(2)}$ and the $\chi^{(3)}$ non-linearities shifts the Hopf 
bifurcation of the standard SHG towards higher intra-cavity energies 
eventually completely stabilizing the system.  Remarkably, the noise
suppression is at the same time strongly enhanced, so that almost perfect
squeezing is obtained for arbitrarily large intra-cavity photon
numbers. Possible experimental implementations are finally discussed.
\end{abstract}

\pacs{PACS numbers: 42.50.Dv, 42.50.Lc, 42.65.Ky}
]

\narrowtext
 From the pioneering ``proof of the principle'' experiment (only 0.3 dB of 
noise reduction) of Slusher and coworkers \cite{Slu85}, the generation of 
squeezed light has been steadily improving.  Thus, in a landmark 
experiment, Schiller and coworkers \cite{Schi96} have very recently got a 
squeezed vacuum with 5.5 dB of noise suppression.  With such a preparation 
they managed to determine completely the state (by measuring the density 
matrix) and observed for the first time the weird oscillations in the 
photon number probability distribution some time ago predicted \cite{Sch87}.  
Paschotta and coworkers \cite{Pas94} have also demonstrated a stable 130 mW cw
source of 1.5 dB squeezed light while Tsuchida, using essentially the
same setup \cite{Tsu95}, has achieved an impressive 65 mW cw source with
a measured 2.4 dB noise suppression (5.2 dB accounting for detection 
efficiency). Similarly, Bergman and coworkers have got a pulsed source
with 5.1 dB of broad band noise reduction \cite{Ber94}.  
Indeed, a figure of 10 dB seems reachable in the near future.

All the above mentioned experiments are based in pure $\chi^{(2)}$ or 
$\chi^{(3)}$ non-linear interactions.  Given the relative simplicity of 
these systems, they have enjoyed the favor of both theoreticians and 
experimentalists.  There were, however, some relatively early incursions 
\cite{Tom84,Tom86} in more complicated systems combining 
both kinds of non-linearities, suggesting enhanced quantum noise reduction.  
Recently, some theoretical work has confirmed this possibility in two 
different experimental configurations \cite{Cab93,Sun95}.  Here, we 
present a third approach in which the competition between the two 
non-linearities is advantageously exploited not only for direct
squeezing enhancement but to control the dynamical 
instabilities (closely related to quantum noise reduction). The
benefits of competing non-linearities are not limited only to systems
with $\chi^{(2)}$ and $\chi^{(3)}$. Thus, higher non-linearities
could be even better for quadrature squeezing \cite{Tom88,Tom89} while 
combining two different kinds of $\chi^{(2)}$ non-linearities improves
the generation of twin beams in non-degenerate systems \cite{Mar95a,Mar95b}. 
  
The system consists in a resonant Second Harmonic Generation 
configuration extended with an intra-cavity Kerr-like nonlinearity.
To specify such a model system, two modes with frequencies 
$\omega$ and $2 \omega$ respectively, both resonant 
in a ring cavity with only one mirror with finite reflectivity, 
interact with a suitable nonlinear medium characterized by its second 
$\chi^{(2)}_{2\omega}$ and third $\chi^{(3)}_{\omega,-\omega,\omega}$ order 
susceptibilities.  An effective interaction Hamiltonian can be 
written in a suitable rotating frame as ($\hbar=1 $)
\begin{equation}
 H_{I}=i \frac{\kappa}{2} (a^{\dagger 2} b -a^{2} b^{\dagger} ) + 
 \frac{\Gamma}{2} a^{\dagger 2} a^{2} \,,
\end{equation}

\noindent
where $\kappa$ and $\Gamma$ are proportional to the $\chi^{(2)}_{2\omega}$ and 
$\chi^{(3)}_{\omega,-\omega,\omega}$ susceptibilities respectively, and
$a$, $b$ are the usual annihilation operators for the fundamental and
the second harmonic mode respectively.  Following the standard procedure first
introduced in ref.\cite{Coll85}, a couple of stochastic differential
equations are obtained using a P representation.  A convenient
normalization is achieved by defining
\[
\alpha  =  \sqrt{\frac{\kappa^{2}}{2\gamma_{a}\gamma_{b}}}\, a
,\;
\beta  =  \frac{\kappa}{\gamma_{a}}\, b \,,
\]

\noindent
being $ \gamma_{a} $ and $ \gamma_{b} $ the loss rates through the 
input-output mirror for the $ \omega $, $ 2\omega$ modes. 
The deterministic part of the stochastic equations then takes the form
\begin{mathletters}
\label{eq:deteqn}
\begin{eqnarray}    
\frac{d\alpha}{d\tau} & = & -\alpha + \beta\alpha^{*} - 
i\Lambda \alpha^{*}\alpha^{2}+\lambda\, , 
\label{eq:deteqna}\\
\frac{1}{r}\frac{d\beta}{d\tau} & = & -\beta - \alpha^{2} \,,
\label{eq:deteqnb}
\end{eqnarray}
\end{mathletters}
    
\noindent
where
\[
\tau\,=\, \gamma_{a}t\,,\:r = \frac{\gamma_{b}}{\gamma_{a}}\,,\: 
\Lambda\,=\,\frac{2\gamma_{b}\Gamma}{\kappa^{2}}\,,\:
\lambda \,=\,\sqrt{\frac{\kappa^{2}}{\gamma_{a}^{2}\gamma_{b}}}\,a_{in} \,,
\]

\noindent                                                         
being $a_{in}$ the external amplitude of the coherent driving field.

A linearization procedure \cite{Coll85} yields a set of linear
stochastic differential equations which are formally the same than those given 
in a previous work \cite{Cab92}, since only the inhomogeneous part of the
differential equations has been changed. The explicit form for 
the eigenvalues is \cite{Cab92},
\begin{mathletters}
\label{eq:eigen}
\begin{eqnarray}
k_{1},k_{2} & = & \frac{-r-1+g \mp \sqrt{(-r+1-g)^{2} 
-8r|\alpha_{f}|^{2}}}{2} \,, 
\label{eq:eigena} \\
k_{3},k_{4} & = & \frac{-r-1-g \mp \sqrt{(-r+1+g)^{2} 
-8r|\alpha_{f}|^{2}}}{2} \,,
\label{eq:eigenb}
\end{eqnarray}
\end{mathletters}

\noindent
where $g = |\alpha_{f}|^{2}\sqrt{1-3\Lambda^{2}}$ and $\alpha_{f}$
denotes a fixed point of Eqs. (\ref{eq:deteqn}). 
The stability analysis evidences a Hopf bifurcation \cite{Guc86}
at $ g = r + 1 $, where $ {\rm Re}\,k_{1} = {\rm Re}\,k_{2} = 0$ and $
{\rm Im}\,k_{1} = - {\rm Im}\,k_{2} \neq 0 $. 
Self-sustained oscillations build up,
then, above a critical intra-cavity ``classical'' photon number given by
\begin{equation}
\label{nc}
n_{c} \equiv |\alpha_{f}^{c}|^{2} = \frac{r+1}{\sqrt{1-3\Lambda^{2}}}\,.
\end{equation}

\noindent
This critical value reaches infinity at $\Lambda =1/\sqrt{3}$ and indeed
it can be shown that for larger $\Lambda$ the system is stable. 

The efficiencies for each mode (the ratio between the input and
the output powers) can be expressed as
\begin{mathletters}
\label{eq:effi}
\begin{eqnarray}
\eta_{a} & = & 
\frac{(1-n)^{2}+n^{2}\Lambda^{2}}{(1+n)^{2}+n^{2}\Lambda^{2}} \,,
\label{eq:effia}
\\
\eta_{b} & = & \frac{4n}{(1+n)^{2}+\Lambda^{2}n^{2}} \,,
\label{eq:effib}
\end{eqnarray}
\end{mathletters}

\noindent
where we have defined $n \equiv |\alpha_{f}|^{2}$. As expected, the
$\chi^{(3)}$ non-linearity diminishes the SHG efficiency (from here the
term ``competing''). Although at a first glance this appears as a
disappointing consequence, from the noise reduction perspective, 
it will turn out very advantageous.

When no Kerr effect is present, perfect squeezing can be reached at the 
critical point in the limits of total asymmetry in the losses \cite{Coll85}
($r\rightarrow 0$ for perfect squeezing in the fundamental, 
$r \rightarrow \infty$ for the harmonic mode).
We proceed now, studying how the squeezing is affected by the third 
order non-linearity.  Quantum noise depends on the phase of the chosen 
quadrature as well as on the frequency.  The squeezing spectrum can be 
analytically optimized on the quadrature phase resulting for the maximally 
squeezed and the corresponding conjugate quadrature 
(see for example \cite{Fab90})
\begin{eqnarray}
\label{eq:spec}
S_{-,+}(\omega) & = & N \int_{-\infty}^{\infty} \{ \langle 
a_{out}^{\dagger}(t+\tau),a_{out}(t) \rangle \mp \nonumber \\
& &
\left | \langle:a_{out}(t+\tau),a_{out}(t): \rangle \right | \} \, 
e^{i\omega \tau}\: d \tau \,,
\end{eqnarray}

\noindent
where $a_{out}$ is the annihilation operator of the corresponding mode 
for the out-going field, colons denote normal order and $N$ is a 
normalization constant.  With this normal order definition, 
$S_{-,+}(\omega)$ is proportional to the intra-cavity normal and time ordered  
spectrum with proportionality constant the cavity losses 
\cite{Coll84,Gar85}.  The maximum squeezing available for fixed values of 
$n$, $\Lambda$ and $r$ (denoted by $S_{-}(\omega_{m})$) is obtained by 
numerical optimization of $S_{-}(\omega)$ with respect to the frequency.  

\begin{figure}[btp]
\centerline{\psfig{figure=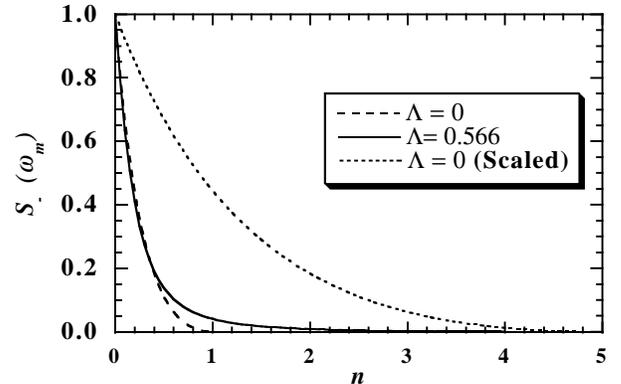,width=8.5cm,bbllx=2.2cm,bblly=11.6cm,bburx=16.3cm,bbury=20.5cm,clip=}}
\caption{The maximum squeezing in the fundamental mode in Vacuum Noise
Units (i.e., the Vacuum Noise equals one) as a function of
the photon-number for the ideal case $r = 10^{-6}$. $\Lambda = 0.566$
corresponds to $n_{c} = 5$. The scaled SHG corresponds to a coupling
constant $\kappa^{\prime} = \kappa/5$.}
\label{fg:perfect}
\end{figure}

We first center on an almost ideal case with $r = 10^{-6}$.  As mentioned 
above, the fundamental mode noise is then highly suppressed at the critical 
point and therefore is a demanding reference point. Figure~\ref{fg:perfect}
shows $S_{-}(\omega_{m})$ for two pure SHG cases and a $\Lambda$
corresponding to $n_{c} = 5$. The scaled SHG case corresponds to a shift
of the critical point up to $n_{c} = 5$ by a reduction of the coupling 
constant, $\kappa$, induced, for example, by a slight phase mismatch. It
is the natural reference case for squeezing efficiency comparison.
The result can hardly be better.  The $\chi^{(3)}$ non-linearity not
only shifts the critical point but strongly enhances the noise 
suppression performance. The squeezing is even slightly better than the 
$\Lambda = 0$ case for low $n$ (of course, it cannot beat it around
$n=1$ where essentially perfect squeezing is reached for $\Lambda =0$).

\begin{figure}[tbp]
\centerline{\psfig{figure=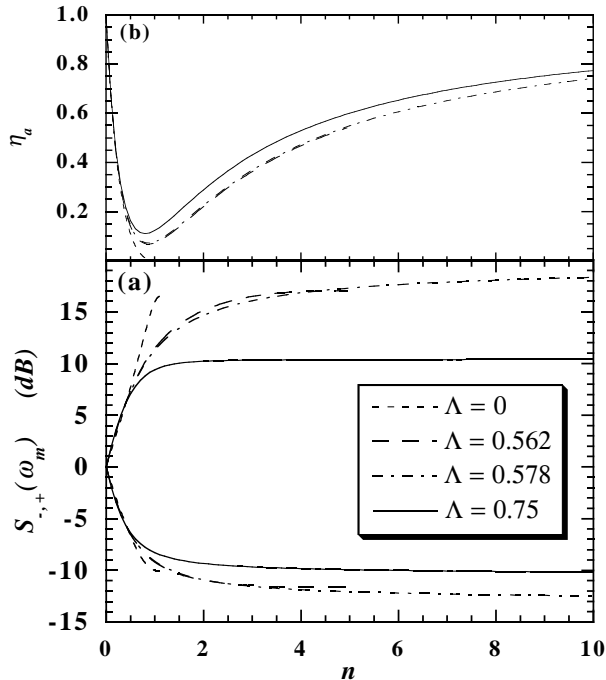,width=8.5cm,bbllx=1cm,bblly=8.4cm,bburx=16.5cm,bbury=25.3cm,clip=}}
\caption{The maximum noise reduction and the excess noise in the
conjugate quadrature (fundamental mode) for $r = 0.15$ {\bf (a)} and the
corresponding $\eta_{a}$ efficiency {\bf (b)}; $\Lambda = 0.562$ implies
$n_{c} = 5$, $\Lambda = 0.578$ $n_{c} = \infty$ and $\Lambda = 0.75$ was
chosen to equal the pure SHG case but at $n = 10$.}
\label{fg:real}
\end{figure}

Turning now to a realistic case we chose a mild asymmetry of 
$r = 0.15$ which gives for the standard SHG a good 10 dB of maximum noise 
reduction.  The behavior of the system is summarized in
figure~\ref{fg:real}, now showing both noise in the squeezed and conjugate 
quadrature (this time in dB's relative to the vacuum noise), as well as the 
``efficiency'' $\eta_{a}$.  As it is shown, the squeezing even increases when 
$n_{c}$ is shifted while the noise in the conjugate quadrature remains 
bounded, slightly surpassing the SHG case.  The case $\Lambda = 0.75$ was 
chosen to obtain the 10 dB around $n = 10$.  Interestingly enough, $\Lambda 
= 0.75$ is above the critical value $1/\sqrt{3}$, therefore being always 
stable.  Notice that the excess noise in the conjugate quadrature nearly 
equals the noise suppression in the squeezed one, i.e., the limit imposed by 
the Heisenberg principle.  With respect to the efficiency, $\eta_{a}$ 
increases above a certain minimum and for the cases shown, it reaches
around 75 \% in the range of the figure.  In other words, when the 
system is thought as a squeezer instead as a frequency doubler, the 
expressions (\ref{eq:effi}) are good news, specially when compared with the 
pure SHG system for which at the point of maximum squeezing the output is 
zero.  That is the reason why we have centered the analysis in the 
fundamental mode.  However, although $\eta_{b}$ decreases with $n$, the 
squeezing behavior of the harmonic mode for $r > 1$ is as excellent as the 
one shown for the fundamental.

\begin{figure}[tbp]
\centerline{\psfig{figure=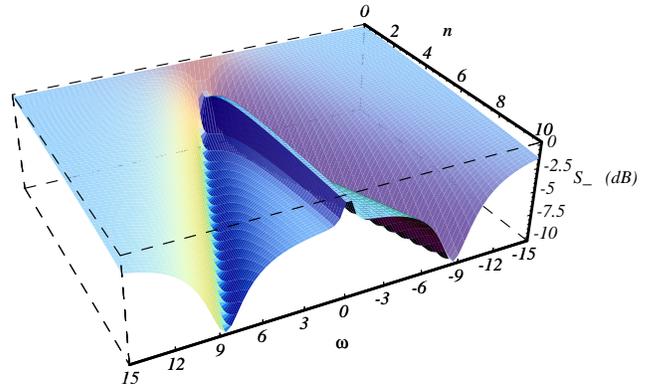,width=8.5cm,bbllx=3.4cm,bblly=16.9cm,bburx=18.3cm,bbury=27.1cm,clip=}}
\caption{The optimized squeezing spectrum (fundamental mode) as a function of
the photon-number for $r = 0.15$ and $\Lambda = 0.75$.}
\label{fg:spec}
\end{figure}

Figure~\ref{fg:spec} displays another feature of the system, namely, the
frequency at which the noise reduction is maximum grows up linearly with $n$.
The third order non-linear term in Eq. (\ref{eq:deteqna}) is the cause of
this frequency shift, since it can be interpreted as a non-linear
detuning of value $\Lambda n$. Such an effect has a twofold benefit.  
From one hand maximum squeezing is not located at zero
frequency, where balanced homodyne detection does not work properly.  
On the other, it can be used as a simple mechanism for tuning the squeezing.
Quite a similar behavior appears also in the non-degenerate case studied in
\cite{Mar95a,Mar95b}. 

The most obvious way of practical implementation of the system
would be to place two different materials in the same cavity, one suited
for the second order interaction, the other for the third. However,
for the low values of $\Lambda$ needed, 
a simple crystal could be adequate for the task. For example, from the KDP
electrooptic Kerr coefficients ($s_{66} = 0.9 \times 10^{-18}\,
{\rm m^{2}/V^{2}}$, \cite{Agu94}), a $\chi^{(3)}_{0,0,\omega}$ of 
$1.5 \times 10^{-18}\,{\rm m^{2}/V^{2}}$ is estimated. To take
into account possible non-electronic contributions (KDP is a
ferroelectric, see \cite{Agu94}) we
shall assume a $\chi^{(3)}_{\omega,-\omega,\omega}$ in the range
$1.5 \times 10^{-19}-10^{-18}$. From the definitions of $\kappa$, $\Gamma$ and
$\gamma_{b}$ and assuming plane wave modes in the resonator, it can be
shown that $\Lambda =
(T_{b}/4\pi)(\lambda_{b}/l)(\chi^{(3)}/(\chi^{(2)})^{2})$ being $T_{b}$
and $\lambda_{b}$ the transmission at the mirror and the wavelength for
mode $b$ and $l$ the crystal length. Taken 
$\chi^{(2)}_{2 \omega} = 10^{-12}\, {\rm V/m}$ \cite{Agu94}, 
$T_{b} = 0.2$, $\lambda_{b}= 1.06$ $\mu$m and $l =$ 1 cm, 
the range for $\Lambda$ would be $0.23-2.3$, suitable for
our purposes. At a first glance, $\Lambda$ can be easily increased by
shortening the crystal length. However, shorter lengths imply larger
input powers to reach the pure SHG critical point. As the
interesting behavior is above this point, power requirements limit this
possibility. Fortunately, these requirements are not very demanding
in a completely resonant configuration (resonant for all the
interacting modes). For example, in \cite{Ric95} a
triply resonant OPO is achieved showing a threshold power below 1
mW. Stable double resonance is possibly the most difficult experimental
task, but the above mentioned experiment \cite{Ric95}, as well as the
results in \cite{Kur93}, certainly demonstrated that it is possible.

A second very interesting approach is the use of Asymmetric Multiple Quantum
Wells. These devices are currently object of an intense study regarding its
applicability for SHG \cite{Sir92,Jan94,Mil96,Vur96}. For our purposes,
they are excellent candidates as the energy levels schemes used for 
$\chi^{(2)}$ optimization are, at the same time, adequate for $\chi^{(3)}$
enhancement. Even more, the non-linearities can be controlled with an
applied dc field \cite{Sir92,Vur96}. A drawback for such systems is
the increase of two-photon absorption associated with the resonant
enhancement of non-linearities. Although not necessarily pernicious for
the squeezing (two photon absorption is also a source of squeezing), the 
augmented total losses could force to use pulsed light so that the results
shown here are not directly applicable. However, as we are not interested in a
giant $\chi^{(2)}$ but in a compensated ratio between $\chi^{(3)}$ and
$\chi^{(2)}$, resonance is no so critical and maybe an adequate working
point with a sufficiently diminished two-photon absorption could be found.

Cascaded second-order nonlinearities (see for instance \cite{Whi96} for
an experimental demonstration of dispersive bistability by cascading
effect) are also very interesting since the $\chi^{(2)}$ and the 
effective $\chi^{(3)}$ are, as in the previous cases, naturally embedded in
the same material. Unfortunately, complete resonance must be abandoned
in this case and again the present results need revision. This
possibility is currently under investigation by the authors. 

Finally, poled fibers \cite{Kaz94} which generate a SHG signal would be a 
technologically very attractive possibility. Of course, pulses are
needed in such a case in order to achieve an appreciable self-phase
modulation.    

To conclude, a summary of the properties of our model system is in order. 
First, arbitrarily large squeezing is obtainable at arbitrarily large output
intensities. Second, both the squeezing and the efficiency improve with
increasing output intensity. Third, the excess noise in the conjugate
quadrature grows up moderately. Fourthly, the maximum squeezing is not
at zero frequency. As a by-product, the squeezing can be linearly tuned
by control of the input intensity. Good figures are obtained for physically
reasonable parameters (figure \ref{fg:real}) and the experimental
implementation seems feasible. Altogether make the proposed model
system an excellent candidate for bright squeezing generation.

Work supported in part by project No. TIC95-0563-C05-03 (C.I.C.Y.T., Spain).
C. C. wishes to thank S. Schiller, C. Fabre and F. Agull{\'o}-L{\'o}pez for useful 
conversations and R. Serna for providing valuable bibliography and
non-linear materials data.


\begin{thebibliography}{99}

\bibitem{Slu85} R. E. Slusher et al, Phys. Rev. Lett. 55, 2409 (1985).

\bibitem{Schi96} S. Schiller et al, Phys. Rev. Lett. 77, 2933 (1996).

\bibitem{Sch87}  W. Schleich and J. A. Wheeler, Nature 326, 574 (1987).

\bibitem{Pas94} R. Paschotta et al, Phys. Rev. Lett. 72, 3807 (1994).

\bibitem{Tsu95} H. Tsuchida, Opt. Lett. 20, 2240 (1995)

\bibitem{Ber94} K.  Bergman et al,  Opt. Lett. 19, 290 (1994)

\bibitem{Tom84} P. Tombesi and H. P. Yuen in {\it Coherence and Quantum
Optics V}, edited by L. Mandel and E. Wolf (Plenum, New York, 1984).

\bibitem{Tom86} P. Tombesi, in {\it Quantum Optics IV}, Proceedings of
the Fourth International Symposium, Hamilton, new Zealand, 1986, edited
by J. D. Harvey and D. F. Walls, (Springer--Verlag, Berlin,1986).

\bibitem{Cab93} C. Cabrillo and F. J. Bermejo, Phys. Rev. A 48, 2433 (1993).

\bibitem{Sun95} K. Sundar, Phys. Rev. Lett. 75, 2116 (1995).

\bibitem{Tom88} P. Tombesi and A. Mecozzi, Phys. Rev. A 37, 4778 (1988).

\bibitem{Tom89} P. Tombesi, Phys. Rev. A 39, 4288 (1989)

\bibitem{Mar95a} M. A. M. Marte, Phys. Rev. Lett. 76, 4815 (1995)

\bibitem{Mar95b} M. A. M. Marte, J. Opt. Soc. Am. B 12, 2296 (1995)

\bibitem{Coll85} M. J. Collet and D. F. Walls, Phys. Rev. A 32, 2887
(1985).

\bibitem{Cab92} C. Cabrillo and F. J. Bermejo, Phys. Lett. A 170, 300 (1992).

\bibitem{Guc86} J. Guckenheimer and P. Holmes, {\it Nonlinear Oscillations,
Dynamical Systems, and Bifurcations of Vector Fields}
(Springer-Verlag, New York, 1986).

\bibitem{Fab90} C. Fabre et al, Quantum Opt. 2, 159 (1990).

\bibitem{Coll84} M. J. Collet and C. W. Gardiner, Phys. Rev. A 30, 1386 (1984)

\bibitem{Gar85} C. W. Gardiner and M. J. Collet, Phys. Rev. A 31, 3761
(1985).

\bibitem{Agu94} F. Agull{\'o}-L{\'o}pez, J. M. Cabrera, F. Agull{\'o}-Rueda,
{\it Electrooptics: Phenomena, Materials and applications} (Academic
Press, San Diego, 1994).

\bibitem{Ric95} C. Richy et al, J. Opt. Soc. Am. B 12, 456 (1995).

\bibitem{Kur93} P. Kurz et al, Europhys. Lett. 24, 449 (1993). 

\bibitem{Sir92} C. Sirtori et al, Appl. Phys. Lett. 60, 151 (1992).

\bibitem{Jan94} S. Janz, F. Chatenoud, and R. Normandin, Opt. Lett. 19,
622 (1994).

\bibitem{Mil96} V. Milanovic and Z. Ikonic, IEEE J. Quan. Elect. 32,
1316 (1996).

\bibitem{Vur96} I. Vurgaftman et al, IEEE J. Quan. Elect. 32, 1334
(1996).

\bibitem{Whi96} A. G. White, J. Mlynek and S. Schiller,
Europhys. Lett. 35, 425 (1996).

\bibitem{Kaz94} P. G. Kazansky, L. Dong, and P. St. J. Russell,
Opt. Lett. 19, 701 (1994).

\end{thebibliography}
\end{document}